\documentclass{article}%
\usepackage{amsfonts}
\usepackage{amsmath}
\usepackage{amssymb}
\usepackage{graphicx}%
\providecommand{\U}[1]{\protect\rule{.1in}{.1in}}
\providecommand{\U}[1]{\protect\rule{.1in}{.1in}}
\newtheorem{theorem}{Theorem}

\newtheorem{corollary}[theorem]{Corollary}

\newtheorem{definition}[theorem]{Definition}
\newtheorem{example}[theorem]{Example}

\newtheorem{lemma}[theorem]{Lemma}

\newtheorem{proposition}[theorem]{Proposition}
\newtheorem{remark}[theorem]{Remark}

\newenvironment{proof}[1][Proof]{\noindent\textbf{#1.} }{\ \rule{0.5em}{0.5em}}
\begin{document}

\title{On reciprocal equivalence of St\"{a}ckel systems}
\author{Maciej B\l aszak\\Faculty of Physics, Division of Mathematical Physics, A. Mickiewicz University\\Umultowska 85, 61-614 Pozna\'{n}, Poland\\blaszakm@amu.edu.pl
\and Krzysztof Marciniak\\Department of Science and Technology \\Campus Norrk\"{o}ping, Link\"{o}ping University\\601-74 Norrk\"{o}ping, Sweden\\krzma@itn.liu.se}
\maketitle

\begin{abstract}
In this paper we ivestigate St\"{a}ckel transforms between different classes
of parameter-dependent St\"{a}ckel separable systems of the same dimension. We
show that the set of all St\"{a}ckel systems of the same dimension splits to
equivalence classes so that all members within the same class can be connected
by a single St\"{a}ckel transform. We also give an explicit formula relating
solutions of two St\"{a}ckel-related systems. These results show in particular
that any two geodesic St\"{a}ckel systems are St\"{a}ckel equivalent in the
sense that it is possible to transform one into another by a single
St\"{a}ckel transform. We also simplify proofs of some known statements about
multiparameter St\"{a}ckel transform.

\end{abstract}

Keywords and phrases: Hamiltonian systems, completely integrable systems,
St\"{a}ckel systems, Hamilton-Jacobi theory, St\"{a}ckel transform

\section{Introduction}

St\"{a}ckel transform is a functional transform that transforms a given
Liouville integrable system into a new integrable system on the same Poisson
manifold. It was first described by J. Hietarinta et al in \cite{hietarinta}
(where it was called the coupling-constant metamorphosis) and developed in
\cite{boyeretal}. It has been applied in \cite{kkmw, kkmw2,kkmw3,kkmw4} for
the purpose of classification of superintegrable systems in conformally flat
spaces. In \cite{ts,ts2} the author described St\"{a}ckel transform as a
canonical transformation on an extended phase space. Applied to a St\"{a}ckel
separable system, this transformation yields a new St\"{a}ckel separable
system, which explains its name.

Originally, only one coupling constant, entering linearly in one of the
Hamiltonians of the system, was used. In paper \cite{macartur2008} a
multiparameter generalization of St\"{a}ckel transform has been introduced.
This generalization allows for a nonlinear dependence of Hamiltonians of the
system on several coupling parameters, thus much enlarging the class of
admissible St\"{a}ckel transforms. Also, this generalized transform results in
a class of reciprocal transformations that has been applied in
\cite{macartur2009} for analyzing weakly-nonlinear semi-Hamiltonian
hydrodynamic-type systems. This indicates that St\"{a}ckel transform is a
useful tool for studying various integrable systems. It can also be
generalized for studying systems of ODE's of evolutionary type with integrals
of motion, see \cite{arturODE}.

In this paper we use the approach developed in \cite{macartur2008} and further
in \cite{macartur2011} to show that all St\"{a}ckel systems of the same
dimension $n$ can be split into equivalence classes such that every two
members of the same class \ are St\"{a}ckel equivalent in the sense that there
always exists a single $n$-parameter St\"{a}ckel transform between arbitrary
two such systems. In order to do this we consider St\"{a}ckel transforms
inside given classes of St\"{a}ckel systems, a problem not considered in
previous papers. We also give an explicit, compact form of this transform,
making the formulas more transparent then these in \cite{macartur2008} and
\cite{macartur2011}. Also, we present a corresponding reciprocal transform
between solutions of these St\"{a}ckel-related systems confined to proper
submanifolds of the phase space. We also clarify and in some cases also repair
a number of formulas and simplify proofs of a number of statements given in
\cite{macartur2008}. Two extensive examples are given at the end of the paper.

\section{General St\"{a}ckel transform}

In this section we present some facts about multiparameter St\"{a}ckel
transform. Consider a manifold $M$ equipped with a Poisson tensor $\Pi$.
Denote the space of all smooth functions on $M$ by $C^{\infty}(M)$. The
mapping $\left\{  \cdot,\cdot\right\}  :C^{\infty}(M)\times C^{\infty
}(M)\rightarrow C^{\infty}(M)$ given by $\left\{  f,g\right\}  _{\Pi}=(df,\Pi
dg)$ (where $\left(  \cdot,\cdot\right)  $ is the dual map between cotangent
and tangent spaces) is called Poisson bracket and it turns $C^{\infty}(M)$
into a Lie algebra. Suppose we have $r$ functions (later: Hamiltonians)
$h_{i}:M\rightarrow R$ on $M,$ each depending on $k\leq r$ parameters
$\alpha_{1},\dots,\alpha_{k}$ so that
\begin{equation}
h_{i}=h_{i}(x,\alpha_{1},\dots,\alpha_{k}),\quad i=1,\dots,r, \label{hi}%
\end{equation}
where $x\in M$. Let us now from $r$ functions in (\ref{hi}) choose $k$
functions $h_{s_{i}}$, $i=1,\ldots,k$, where $\{s_{1},\dots,s_{k}\}$
$\subset\{1,\dots,r\}$. Assume also that the system of equations%
\begin{equation}
h_{s_{i}}(x,\alpha_{1},\dots,\alpha_{k})=\tilde{\alpha}_{i},\quad i=1,\dots,k,
\label{hsi}%
\end{equation}
(where $\tilde{\alpha}_{i}$ is another set of $k$ free parameters, or values
of Hamiltonians $h_{s_{i}}$) involving the functions $h_{s_{i}}$ can be solved
for the parameters $\alpha_{i}$ yielding%
\begin{equation}
\alpha_{i}=\widetilde{h}_{s_{i}}(x,\tilde{\alpha}_{1},\dots,\tilde{\alpha}%
_{k}),\quad i=1,\dots,k, \label{alphy}%
\end{equation}
where the right hand sides of these solutions define $k$ new functions
$\widetilde{h}_{s_{i}}$ on $M$, each depending on $k$ parameters
$\tilde{\alpha}_{i}$. Finally, let us define $r-k$ functions $\widetilde
{h}_{i}$ with $i=1,\ldots,r$ and such that $i\notin\{s_{1},\dots,s_{k}\}$ by -
in accordance with (\ref{alphy}) - substituting $\widetilde{h}_{s_{i}}$
instead of $\alpha_{i}$ in $h_{i}$ for $i\notin\{s_{1},\dots,s_{k}\}$:%
\begin{equation}
\widetilde{h}_{i}=h_{i}|_{\alpha_{1}\rightarrow\widetilde{h}_{s_{1}}%
,\ldots,\alpha_{k}\rightarrow\widetilde{h}_{s_{k}}},\quad i=1,\dots,r,\text{
\ }i\notin\{s_{1},\dots,s_{k}\}. \label{reszta}%
\end{equation}

\begin{definition}
The functions $\widetilde{h}_{i}=\widetilde{h}_{i}(x,\tilde{\alpha}_{1}%
,\dots,\tilde{\alpha}_{k}),$ $i=1,\dots,r$, defined through (\ref{alphy}) and
(\ref{reszta}) are called the (generalized) St\"{a}ckel transform of the
functions (\ref{hi}) with respect to the indices $\{s_{1},\dots,s_{k}\}$ (or
with respect to the functions $h_{s_{1}},\ldots h_{s_{k}}$).
\end{definition}

Note that unless we extend the manifold $M$ this operation can in general not
be obtained by any coordinate change of variables. It is also easy to see that
if we perform again the St\"{a}ckel transform on the functions $\widetilde
{h}_{i}$ with respect to $\widetilde{h}_{s_{i}}$ we will receive back the
functions $h_{i}$ in (\ref{hi}). Note also that neither $k$ nor $r$ are
related to the dimension of the manifold $M$.

\begin{example}
The simplest situation occurs when $k=r=1$. Consider, after \cite{hietarinta},
the Fokas-Lagerstr\"{o}m potential on the four-dimensional phase space $M$
with coordinates $(x,y,p_{x},p_{y}):$%
\[
h=\frac{1}{2}(p_{x}^{2}+p_{y}^{2})-\frac{2}{3}\alpha(xy)^{-2/3}%
\]
Solving the equation $h=$ $\tilde{\alpha}$ with respect to the only parameter
$\alpha$ (called in \cite{hietarinta} a coupling constant) one obtains%
\[
\alpha=\frac{3}{4}(xy)^{2/3}(p_{x}^{2}+p_{y}^{2})-\frac{3}{2}(xy)^{2/3}%
\widetilde{\alpha}\equiv\widetilde{h}%
\]
which can be shown \cite{hietarinta} to be equivalent to the axially symmetric
potential $\rho^{4}$.
\end{example}

St\"{a}ckel transform has two important properties that make it well suited
for study of integrable systems: as we will see in Theorem \ref{zachowuje}, it
preserves functional independence and it also preserves involutivity with
respect to the Poisson tensor $\Pi$. Moreover, as it will also be demonstrated
in this paper, it maps a St\"{a}ckel separable system into a new St\"{a}ckel
separable system which explains the name of this transformation.

In the special but nonetheless important for this paper case when functions
(\ref{hi}) depend linearly on parameters $\alpha_{i}$ it is possible to write
down the St\"{a}ckel transform explicitly. Suppose therefore for the moment
that the functions in (\ref{hi}) have the form%

\begin{equation}
h_{i}=H_{i}+\sum\limits_{j=1}^{k}\alpha_{j}H_{i}^{(j)},\quad i=1,\dots,r.
\label{lin}%
\end{equation}
The equations (\ref{hsi}) defining the first part of the St\"{a}ckel transform
take then the form of a system of $k$ linear equations in $k$ unknowns
$\alpha_{1},\ldots,\alpha_{k}$
\[
H_{s_{i}}+\sum\limits_{j=1}^{k}\alpha_{j}H_{s_{i}}^{(j)}=\tilde{\alpha}%
_{i},\quad i=1,\dots,k,
\]
with the Cramer solution for $\alpha_{i}=\widetilde{h}_{s_{i}}$ of the form:
\begin{equation}
\widetilde{h}_{s_{i}}=\det W_{i}/\det W, \label{detlin}%
\end{equation}
where
\[
W=\left\vert \!%
\begin{array}
[c]{ccc}%
H_{s_{1}}^{(1)} & \cdots & H_{s_{1}}^{(k)}\\
\vdots & \ddots & \vdots\\
H_{s_{k}}^{(1)} & \cdots & H_{s_{k}}^{(k)}%
\end{array}
\!\right\vert
\]
is the $k\times k$ matrix $\det\left(  \partial h_{s_{i}}/\partial\alpha
_{j}\right)  $ (so that $\det W\neq0$) and where $W_{i}$ are obtained from $W$
by replacing $H_{s_{j}}^{(i)}$ in the $i$-th column by $\tilde{\alpha}%
_{j}-H_{s_{j}}$ for all $j=1,\dots,k$. The second part of the transformation,
i.e. formulas (\ref{reszta}), reads now
\[
\widetilde{h}_{i}=H_{i}+\sum\limits_{j=1}^{k}\widetilde{h}_{s_{j}}H_{i}%
^{(j)},\text{ \ \ \ }i=1,\dots,r,\text{ \ \ }i\notin\{s_{1},\dots,s_{k}\}
\]
where $\widetilde{h}_{s_{i}}$ are given by (\ref{detlin}). For $k=1$ the above
transformation reproduces the original St\"{a}ckel transform presented in
\cite{hietarinta} and \cite{boyeretal}.

\section{St\"{a}ckel transform for completely integrable systems}

Let us now discuss the St\"{a}ckel transform and the corresponding reciprocal
transform between two Liouville integrable systems. Suppose therefore that
$\dim M=2n$ and that we have exactly $n$ (so that $r=n$ now) functionally
independent functions (Hamiltonians)
\[
h_{i}=h_{i}(x,\alpha_{1},\ldots,\alpha_{k})\text{, \ \ }i=1,\ldots,n
\]
that depend on $k\leq n$ parameters $\alpha_{i}$ and that are for all values
of $\alpha_{i}$ in involution with respect to a nondegenerate Poisson bracket
$\Pi$: $\left\{  h_{i},h_{j}\right\}  _{\Pi}=0$ for all $i,j$. These functions
yield $n$ commuting Hamiltonian systems on $M$:%
\begin{equation}
\frac{dx}{dt_{i}}=\Pi dh_{i}\equiv X_{i}\text{, }i=1,\ldots,n \label{ham}%
\end{equation}
(each depending on $k$ parameters $\alpha_{i}$) so that $X_{i}$ are $n$
commuting Hamiltonian vector fields on $M$. Consider now a new set of $n$
functions (Hamiltonians) $\widetilde{h}_{i}$ obtained from $h_{i}$ by a
St\"{a}ckel transform performed with respect to $h_{s_{1}},\ldots,h_{s_{k}}$.
These functions define a set of Hamiltonian flows on $M$, the vector fields of
which are given by%
\begin{equation}
\frac{dx}{d\widetilde{t}_{i}}=\Pi d\widetilde{h}_{i}\equiv\widetilde{X}%
_{i}\text{, }i=1,\ldots,n \label{hamtil}%
\end{equation}
depending on $k$ parameters $\widetilde{\alpha}_{i}$. We will now consider the
relation between the Hamiltonian systems (\ref{ham}) and (\ref{hamtil}). In
order to study this relation it is important to realize that both systems
(\ref{ham})\ and (\ref{hamtil}) are multiparameter and the relation between
them can thus only be found if one fixes the values of both all $\alpha_{i}$
and all $\widetilde{\alpha}_{i}$ which means that the sought relation can only
exists on the $(2n-k)$-dimensiomal submanifolds $M_{\alpha,\widetilde{\alpha}%
}$ given by (\ref{hsi}):%
\begin{equation}
M_{\alpha,\widetilde{\alpha}}=\left\{  x\in M:h_{s_{i}}(x,\alpha_{1}%
,\dots,\alpha_{k})=\tilde{\alpha}_{i}\text{, \ \ }i=1,\ldots k\right\}
\label{Maa}%
\end{equation}
Note that the surfaces $M_{\alpha,\widetilde{\alpha}}$ depend on the
simultanous choice of $2k$ parameters $\alpha_{i}$ and $\widetilde{\alpha}%
_{i}$ and that its codimension is $k$ (so that $\dim M_{\alpha,\widetilde
{\alpha}}=2n-k\geq n$). Note also that due to the equivalence between
(\ref{hsi}) and (\ref{alphy}) the surfaces $M_{\alpha,\widetilde{\alpha}}$ can
equivalently be defined through%
\begin{equation}
M_{\alpha,\widetilde{\alpha}}=\left\{  x\in M:\widetilde{h}_{s_{i}}%
(x,\tilde{\alpha}_{1},\dots,\tilde{\alpha}_{k})=\alpha_{i}\text{,
\ \ }i=1,\ldots k\right\}  \label{Mab}%
\end{equation}

\begin{remark}
\label{r}Through each point $x$ in $M\,\ $there passes infinitely many
submanifolds $M_{\alpha,\widetilde{\alpha}}$. If we fix the values of all the
parameters $\alpha_{i}$ we can for any $x$ always find some values of the
parameters $\tilde{\alpha}_{i}$ so that $x\in M_{\alpha,\widetilde{\alpha}}$;
and vice versa, if we fix $\widetilde{\alpha}_{i}$, for any given $x$ we can
find $\alpha_{i}$ so that $x\in M_{\alpha,\widetilde{\alpha}}$.
\end{remark}

As it follows from (\ref{hsi}) and (\ref{alphy}) the following identity is
valid on $M$ and for all values of parameters $\widetilde{\alpha}_{i}$:
\begin{equation}
h_{s_{i}}(x,\widetilde{h}_{s_{1}}(x,\widetilde{\alpha}_{1},\ldots
,\widetilde{\alpha}_{n}),\ldots,\widetilde{h}_{s_{k}}(x,\widetilde{\alpha}%
_{1},\ldots,\widetilde{\alpha}_{n}))\equiv\widetilde{\alpha}_{i}\text{,
\ }i=1,\ldots,k \label{id1}%
\end{equation}
Moreover, the second part of the transformation, i.e. formula (\ref{reszta})
can be written as the following identity on $M$ (valid again for all values of
$\widetilde{\alpha}_{i}$):%
\begin{equation}
\widetilde{h}_{i}(x,\widetilde{\alpha}_{1},\ldots,\widetilde{\alpha}%
_{n})\equiv h_{i}(x,\widetilde{h}_{s_{1}}(x,\widetilde{\alpha}_{1}%
,\ldots,\widetilde{\alpha}_{n}),\ldots,\widetilde{h}_{s_{k}}(x,\widetilde
{\alpha}_{1},\ldots,\widetilde{\alpha}_{n})),\text{ \ \ }i=1,\ldots,n\text{,
\ }i\notin\{s_{1},\dots,s_{k}\} \label{id2}%
\end{equation}
Differentiating (\ref{id1}) with respect to $x$ we find that \emph{on each}
$M_{\alpha,\widetilde{\alpha}}$
\begin{equation}
dh_{s_{i}}=-\sum_{j=1}^{k}\frac{\partial h_{s_{i}}}{\partial\alpha_{j}%
}d\widetilde{h}_{s_{j}}\text{, \ \ }i=1,\ldots,k \label{trdh1}%
\end{equation}
while differentiation of (\ref{id2}) gives that on $M_{\alpha,\widetilde
{\alpha}}$ we have
\begin{equation}
dh_{_{i}}=d\widetilde{h}_{i}-\sum_{j=1}^{k}\frac{\partial h_{i}}%
{\partial\alpha_{j}}d\widetilde{h}_{s_{j}}\text{, \ \ }i=1,\ldots,n,\text{
\ }i\notin\{s_{1},\dots,s_{k}\}\text{\ .} \label{trdh2}%
\end{equation}
The transformation (\ref{trdh1})-(\ref{trdh2}) on $M_{\alpha,\widetilde
{\alpha}}$ can be written in a matrix form as
\begin{equation}
dh=Ad\widetilde{h} \label{trdh}%
\end{equation}
where we denote $dh=(dh_{1},\ldots,dh_{n})^{T}$ and $d\widetilde
{h}=(d\widetilde{h}_{1},\ldots,d\widetilde{h}_{n})^{T}$ and where the $n\times
n$ matrix $A$ is given by%
\[
A_{ij}=\delta_{ij}\text{ for }j\notin\{s_{1},\dots,s_{k}\}\text{, }A_{is_{j}%
}=-\frac{\partial h_{i}}{\partial\alpha_{j}}\text{ \ for }j=1,\ldots,k
\]
From the structure of the matrix $A$ it follows that
\[
\det A=\pm\det\left(  \frac{\partial h_{s_{i}}}{\partial\alpha_{j}}\right)
\]
so that $\det A\neq0$ due to our assumptions. Thus, the relation (\ref{trdh})
can be inverted yielding $d\widetilde{h}=A^{-1}dh$. This leads to an important
theorem \cite{macartur2008}, mentioned in Section 2.

\begin{theorem}
\label{zachowuje}1. If the functions $h_{i}$ are functionally independent for
all values of $\alpha_{i}\,$then $\widetilde{h}_{i}$ are functionally
independent for all values of $\widetilde{\alpha}_{i}$. 2. If the functions
$h_{i}$ are for all values of $\alpha_{i}$ in involution with respect to the
Poisson tensor $\Pi$ then the functions $\widetilde{h}_{i}$ are also in
involution with respect to $\Pi$ for all values of $\widetilde{\alpha}_{i}$.
\end{theorem}

\begin{proof}
1. Assume that $h_{i}$ are functionally independent for all values of
$\alpha_{i}$. Consider the differentials $d\widetilde{h}_{i}$ at a given point
$x\in M\,\ $\ and for some arbitrary values of $\widetilde{\alpha}_{i}$. Due
to Remark \ref{r} one can always find valus of $\alpha_{i}$ such that $x\in
M_{\alpha,\widetilde{\alpha}}$. By (\ref{trdh}) and by the fact that $\det
A\neq0$ the differentials $d\widetilde{h}_{i}$ linearly independent at $x$
(since $dh_{i}$ are) and since $x$ ia arbitrary, $\widetilde{h}_{i}$ are
functionally independent on the whole $M$. 2. Assume $\left\{  h_{i}%
,h_{j}\right\}  _{\Pi}=0$ for all $i,j=1,\ldots,n$ and for all values of
$\alpha_{i}$. Then, as in the proof of the first statement, at any $x\in M$ we
can choose an appropriate $M_{\alpha,\widetilde{\alpha}}$ so that (\ref{trdh})
is valid and thus
\begin{align*}
\left\{  \widetilde{h}_{i},\widetilde{h}_{j}\right\}  _{\Pi}  &  =\left(
d\widetilde{h}_{i},\Pi d\widetilde{h}_{j}\right)  =\left(  \sum_{l_{1}=1}%
^{n}\left(  A^{-1}\right)  _{il_{1}}dh_{l_{1}},\Pi\sum_{l_{2}=1}^{n}\left(
A^{-1}\right)  _{jl_{2}}dh_{l_{2}}\right) \\
&  =\sum_{l_{1},l_{2}=1}^{n~}\left(  A^{-1}\right)  _{il_{1}}\left(
A^{-1}\right)  _{jl_{2}}\left(  dh_{l_{1}},\Pi dh_{l_{2}}\right)  =\sum
_{l_{1},l_{2}=1}^{n}\left(  A^{-1}\right)  _{il_{1}}\left(  A^{-1}\right)
_{jl_{2}}\left\{  h_{l_{1}},h_{l_{2}}\right\}  _{\Pi}=0
\end{align*}

\end{proof}

Theorem \ref{zachowuje} implies that the system (\ref{hamtil}) is again
Liouville integrable so that St\"{a}ckel transform maps a Liouville integrable
system into a Liouville integrable system.

Since $X_{i}=\Pi dh_{i}$ and $\widetilde{X}_{i}=\Pi d\widetilde{h}_{i}$ we
obtain from (\ref{trdh1})-(\ref{trdh2}) that the Hamiltonian vector fields
$X_{i}=\Pi dh_{i}$ and $\widetilde{X}_{i}=\Pi d\widetilde{h}_{i}$ are on the
appropriate $M_{\alpha,\widetilde{\alpha}}$ related by the following
transformation%
\begin{align}
X_{s_{i}}  &  =-\sum_{j=1}^{k}\frac{\partial h_{s_{i}}}{\partial\alpha_{j}%
}\widetilde{X}_{s_{j}}\text{, \ \ }i=1,\ldots,k\label{trXsi}\\
X_{_{i}}  &  =\widetilde{X}_{i}-\sum_{j=1}^{k}\frac{\partial h_{i}}%
{\partial\alpha_{j}}\widetilde{X}_{s_{j}}\text{, \ \ }i=1,\ldots,n,\text{
\ }i\notin\{s_{1},\dots,s_{k}\}\text{\ } \label{trXi}%
\end{align}

This means that the hamiltonian vector fields $X_{i}$ and $\widetilde{X}_{i}$
span on each $M_{\alpha,\widetilde{\alpha}}$ the same $n$-dimensional
distribution and also that the vector fields $X_{s_{i}}$ and $\widetilde
{X}_{s_{i}}$ span on each $M_{\alpha,\widetilde{\alpha}}$ the same
$k$-dimensional subdistribution of the above distribution. The transformation
(\ref{trXsi})-(\ref{trXi}) on $M_{\alpha,\widetilde{\alpha}}$ can be written
in matrix form as
\begin{equation}
X=A\widetilde{X} \label{trvec}%
\end{equation}
where we denote $X=(X_{1},\ldots,X_{n})^{T}$ and $\widetilde{X}=(\widetilde
{X}_{1},\ldots,\widetilde{X}_{n})^{T}$ and where the $n\times n$ matrix $A$ is
given above.

All the vector fields $X_{i}$ and $\widetilde{X}_{i}$ are naturally tangent
\ to the corresponding $M_{\alpha,\widetilde{\alpha}}$ so that if $x_{0}\in$
$M_{\alpha,\widetilde{\alpha}}$ then the multiparameter (simultaneous)
solution%
\begin{equation}
x=x(t_{1},\ldots,t_{n},x_{0}) \label{msol}%
\end{equation}
of all equations in (\ref{ham}) starting at $x_{0}$ for $t=0$, will always
remain in $M_{\alpha,\widetilde{\alpha}}$ and the same is also true for
multiparameter solutions of (\ref{hamtil}).

The relations (\ref{trXsi})-(\ref{trXi}) can be reformulated in the dual
language, that of reciprocal (multi-time) transformations.

\begin{theorem}
\label{recipth}The reciprocal transformation $\widetilde{t}_{i}=\widetilde
{t}_{i}(t_{1},\ldots,t_{n},x)$ $i=1,\ldots,n$ given on $M_{\alpha
,\widetilde{\alpha}}$ by%
\begin{equation}
d\tilde{t}=A^{T}dt \label{recip}%
\end{equation}
(where $dt=(dt_{1},\ldots,dt_{n})^{T}$ and $d\widetilde{t}=(d\widetilde{t}%
_{1},\ldots,d\widetilde{t}_{n})^{T}$) transforms the $n$-parameter solutions
(\ref{msol}) of the system (\ref{ham}) to the $n$-parameter solutions
$\widetilde{x}=\widetilde{x}(\widetilde{t}_{1},\ldots,\widetilde{t}_{n}%
,x_{0})$ of the system (\ref{hamtil}) (with the same initial condition
$x(0)=x_{0}\in M_{\alpha,\widetilde{\alpha}}$) in the sense that for any
$x_{0}\in M_{\alpha,\widetilde{\alpha}}$ we have%
\[
\widetilde{x}(\widetilde{t}_{1}(t_{1},\ldots,t_{n},x_{0}),\ldots,\widetilde
{t}_{n}(t_{1},\ldots,t_{n},x_{0}),x_{0})=x(t_{1},\ldots,t_{n},x_{0})
\]
for all values of $t_{i}$ sufficiently close to zero.
\end{theorem}

The transformation (\ref{recip}) is well defined since the right hand side of
(\ref{recip}) is an exact differential, as it follows from the above
construction. It means that it is possible (at least locally) to integrate
(\ref{recip}) and obtain an explicit transformation $\tilde{t}_{i}=\tilde
{t}_{i}(t_{1},\ldots,t_{n},x)$ that takes multi-time (simultaneous) solutions
of all hamiltonian systems (\ref{ham}) to multi-time solutions of all the
systems in (\ref{hamtil}).

In a specific but important for us case when $k=n$ (i.e. when the number of
parameters and the number of hamiltonians coincide so that the St\"{a}ckel
transform consist only of the first part i.e. (\ref{alphy})), the matrix $A$
simplifies to
\[
A_{ij}=-\frac{\partial h_{i}}{\partial\alpha_{j}}\text{, \ }i,j=1,\ldots,n
\]
so that the formulas (\ref{trXsi})-(\ref{trXi}) simplify to the formula%
\[
X_{i}=-\sum\limits_{j=1}^{n}\frac{\partial h_{i}}{\partial\alpha_{j}%
}\widetilde{X}_{j},\quad i=1,\dots,n,
\]
while (\ref{recip}) can be explicitly written as%
\begin{equation}
d\tilde{t}_{i}=-\sum\limits_{j=1}^{n}\frac{\partial h_{j}}{\partial\alpha_{i}%
}dt_{j},\quad i=1,\dots,n. \label{trczas}%
\end{equation}
and our manifolds $M_{\alpha,\widetilde{\alpha}}$ become in this case level
surfaces for all the hamiltonians $h_{i}(x,\alpha)$ and also level surfaces
for all the hamiltonians $\widetilde{h}_{i}(x,\widetilde{\alpha})$.

\section{Classical St\"{a}ckel systems}

Consider a set of Darboux coordinates $(\lambda,\mu)=(\lambda_{1}%
\ldots,,\lambda_{n},\mu_{1},\ldots,\mu_{1})$ on our $2n$-dimensional Poisson
manifold $M$ equipped with a Poisson operator $\Pi$ (so that $\Pi=%
{\textstyle\sum\nolimits_{i<j}}
\frac{\partial}{\partial\lambda_{i}}\wedge\frac{\partial}{\partial\mu_{i}}$).
A classical St\"{a}ckel system is a system of $n$ Hamiltonians $H_{i}$ on $M$
(that originally do not depend on any additional parameters $\alpha$ so that
they can not be a subject of any St\"{a}ckel transformation) originating from
a set of $n$ separation relations \cite{Sklyanin} of the form:
\begin{equation}
\sigma(\lambda_{i})+%
{\displaystyle\sum\limits_{j=1}^{n}}
H_{j}\lambda_{i}^{\gamma_{j}}=f(\lambda_{i})\mu_{i}^{2}\text{, \ \ \ }%
i=1,\ldots,n, \label{Stack}%
\end{equation}
where $f$ and $\sigma$ are arbitrary functions of one argument and where all
$\gamma_{i}\in\mathbf{Z,}$ $i=1,\ldots,n,$ and are such that no two
$\gamma_{i}$ coincide. Thus, a particular St\"{a}ckel system is defined by the
choice of integers $\gamma_{1},\ldots,\gamma_{n}$ and by the choice of
functions $f$ and $\sigma$. Customary one can also treat this system of
relations as $n$ points on ($n$ copies of) the following \emph{separation
curve}%
\begin{equation}
P(\lambda,H)\equiv\sigma(\lambda)+%
{\displaystyle\sum\limits_{j=1}^{n}}
H_{j}\lambda^{\gamma_{j}}=f(\lambda)\mu^{2}, \label{Stackc}%
\end{equation}
in $\lambda\mu$ plane which helps us to avoid writing too many indices. The
relations (\ref{Stack}) (or $n$ copies of (\ref{Stackc})) constitute a system
of $n$ equations linear in the unknowns $H_{i}$. Solving these relations with
respect to $H_{i}$ we obtain $n$ commuting (since the right-hand sides of
formulas (\ref{Stack}) commute) with respect to $\Pi$ Hamiltonians (known in
literature as St\"{a}ckel Hamiltonians) on $M$ of the form
\begin{equation}
H_{i}=\mu^{T}K_{i}G\mu+V_{i}(\lambda)\text{ \ }i=1,\ldots,n\text{,}
\label{Stackham}%
\end{equation}
where we denote $\lambda=(\lambda_{1},\ldots,\lambda_{n})^{T}$ and $\mu
=(\mu_{1},\ldots,\mu_{n})^{T}$. The functions $H_{i}$ can be interpreted as
$n$ quadratic in momenta $\mu$ hamiltonians on the phase space $M=T^{\ast
}\mathcal{Q}$ cotangent to a Riemannian manifold $\mathcal{Q}$ (so that
$\lambda_{1,}\ldots,\lambda_{n}$ are coordinates on $\mathcal{Q}$) equipped
with the contravariant metric tensor $G$ depending on function $f$ \ and the
choice of the set $\gamma$. The objects $K_{i}$ in (\ref{Stackham}) can be
interpreted as $(1,1)$-type Killing tensors on $\mathcal{Q}$ for the metric
$G$. The metric tensor $G$ and all the Killing tensors $K_{i}$ \ are diagonal
in $\lambda$-variables. Note that by the very construction of $H_{i}$ the
variables $\left(  \lambda,\mu\right)  $ are separation variables for all the
hamiltonians in (\ref{Stackham}) in the sense that the Hamilton-Jacobi
equations associated with all $H_{i}=a_{i}$ admit additively separable
solutions $W=%
{\textstyle\sum_{i=1}^{n}}
W_{i}(\lambda_{i},a)$.

The relations (\ref{Stack}) can be written in a matrix form as
\[
S_{\gamma}H=U
\]
where $H=(H_{1},\ldots,H_{n})^{T}$, and where $U$ is a St\"{a}ckel vector of
the form%
\begin{equation}
U=(f(\lambda_{1})\mu_{1}^{2}-\sigma(\lambda_{1}),\ldots,f(\lambda_{n})\mu
_{n}^{2}-\sigma(\lambda_{n}))^{T}, \label{Kvec}%
\end{equation}
while the matrix $S_{\gamma}$ is a classical St\"{a}ckel matrix of the form%
\begin{equation}
S_{\gamma}=\left(
\begin{array}
[c]{ccc}%
\lambda_{1}^{\gamma_{1}} & \cdots & \lambda_{1}^{\gamma_{n}}\\
\vdots & \ddots & \vdots\\
\lambda_{n}^{\gamma_{1}} & \cdots & \lambda_{n}^{\gamma_{n}}%
\end{array}
\right)  . \label{Sg}%
\end{equation}
Note that our assumption that no $\gamma_{i}$ coincide means that
$\det(S_{\gamma})\neq0$. Thus, the hamiltonians (\ref{Stackham}) can be
obtained in a matrix form as
\[
H=S_{\gamma}^{-1}U,
\]
which also means that the metric $G$ in (\ref{Stackham}) can be expressed as%
\[
G=\text{diag}\left(  f(\lambda_{1})\left(  S_{\gamma}^{-1}\right)
_{11},\ldots,f(\lambda_{n})\left(  S_{\gamma}^{-1}\right)  _{1n}\right)  ,
\]
so that the Killing tensors $K_{i}$ in (\ref{Stackham}) are%
\[
K_{i}=\text{diag}\left(  \left(  S_{\gamma}^{-1}\right)  _{i1}/\left(
S_{\gamma}^{-1}\right)  _{11},\ldots,\left(  S_{\gamma}^{-1}\right)
_{in}/\left(  S_{\gamma}^{-1}\right)  _{1n}\right)  \text{, \ }i=1,\ldots,n
\]
(note that $K_{1}=I$). Let us now turn our attention to the scalar functions
$V_{i}:\mathcal{Q}\rightarrow\mathbf{R}$ in (\ref{Stackham}). Relations
(\ref{Stackc}) and (\ref{Stackham}) imply that $V_{i}(\lambda)$ satisfy the
following separation curve%
\begin{equation}
\sigma(\lambda)+V_{1}\lambda^{\gamma_{1}}+V_{2}\lambda^{\gamma_{2}}%
+\ldots+V_{n}\lambda^{\gamma_{n}}=0, \label{Vzs}%
\end{equation}
so that they depend on the choice of integers $\gamma_{i}$ and the choice of
the function $\sigma$. We will therefore denote them as $V_{i}^{(\sigma)}$. In
case when $\sigma(\lambda)$ is a monomial, i.e. when $\sigma(\lambda
)=\lambda^{k}$ with $k\in\mathbf{Z}$, $V_{i}$ depend on $k$ and they will be
denoted by $V_{i}^{(k)}$ (so that $V_{i}^{(k)}=V_{i}^{(\lambda^{k})}$) to
shorten the notation. Thus, the potentials $V_{i}^{(k)}(\lambda) $ (they still
depend on all $\gamma_{i}$) satisfy the following separation curve%
\begin{equation}
\lambda^{k}+V_{1}^{(k)}\lambda^{\gamma_{1}}+V_{2}^{(k)}\lambda^{\gamma_{2}%
}+\ldots+V_{n}^{(k)}\lambda^{\gamma_{n}}=0, \label{sepV}%
\end{equation}
which in matrix form can be written as%
\begin{equation}
S_{\gamma}V^{(k)}=-\Lambda^{k}(1,\ldots,1)^{T}, \label{zw}%
\end{equation}
where $V^{(k)}=(V_{1}^{(k)},\ldots,V_{n}^{(k)})^{T}$ and $\Lambda
=$diag$(\lambda_{1},\ldots,\lambda_{n})$. This means that%
\[
V^{(0)}=-S_{\gamma}^{-1}(1,\ldots,1)^{T}%
\]
so that $V^{(1)}=S_{\gamma}^{-1}\Lambda S_{\gamma}V^{(0)}$, $V^{(2)}%
=S_{\gamma}^{-1}\Lambda^{2}S_{\gamma}V^{(0)}=(S_{\gamma}^{-1}\Lambda
S_{\gamma})(S_{\gamma}^{-1}\Lambda S_{\gamma})V^{(0)}$ and so on. Similar
argument applies also for negative $k$. Thus, denoting%
\begin{equation}
F_{\gamma}=S_{\gamma}^{-1}\Lambda S_{\gamma} \label{Fg}%
\end{equation}
we get the compact formula for the potentials $V^{(k)}$ (presented first in
\cite{macartur2011}):
\begin{equation}
V^{(k)}=F_{\gamma}^{k}V^{(0)},\text{ \ \ }k\in\mathbf{Z.} \label{Vk}%
\end{equation}
It is now an immediate consequence of the above formulas that for any
meromorphic function $\sigma(\lambda)$ we have%
\begin{equation}
V^{(\sigma)}=\sigma(F_{\gamma})V^{(0)}, \label{Vsigma}%
\end{equation}
where $V^{(\sigma)}=(V_{1}^{(\sigma)},\ldots,V_{n}^{(\sigma)})$. The matrix
$F_{\gamma}$ given in (\ref{Fg}) has been called control matrix in
\cite{PedroniFalqui} where it appeared in the context of quasi-bi-Hamiltonian
representation. Notice also that if the system (\ref{Stackc}) is normed by
$\gamma_{n}=1,$ then the potential $V^{(0)}$ attains a particularly simple
form $V^{(0)}=(0,\ldots,0,-1)^{T}$. This follows immediately from (\ref{zw}).

Now, by writing down the inverse Jacobi problem for all the Hamiltonians
(\ref{Stackham}) we can arrive at the following remark that will be useful in
the next section when we discuss reciprocal transforms between different
St\"{a}ckel systems.

\begin{remark}
\label{rozw}On the level surface $M_{a}=\left\{  x\in M:H_{i}=a_{i}%
\in\mathbf{R}\right\}  $ the mutliparameter (multi-time) solutions
$\lambda_{i}=\lambda_{i}(t_{1},\ldots,t_{n},x_{0})$ of all Hamiltonian systems
defined by the separation curve (\ref{Stackc}) or equivalently by all
Hamiltonians (\ref{Stackham}) attain the following Abel-Jacobi differential
form
\begin{equation}
dt=S_{\gamma}^{T}\frac{d\lambda}{\sqrt{f(\lambda)P(\lambda,a)}}, \label{sol1}%
\end{equation}
where $d\lambda/\sqrt{f(\lambda)P(\lambda,a)}$ means a column vector with
components $d\lambda_{i}/\sqrt{f(\lambda_{i})P(\lambda_{i},a)}$.
\end{remark}

Note that solutions (\ref{sol1}) define in a standard (canonical) way the
corresponding multi-time solutions for the momenta $\mu_{i}=\mu_{i}%
(t_{1,}\ldots,t_{n},x_{0})$.

A particular subclass of St\"{a}ckel systems is given by choosing $\gamma
_{i}=n-i$. The separation curve (\ref{Stackc}) attains then the form
\[
\sigma(\lambda)+%
{\displaystyle\sum\limits_{j=1}^{n}}
H_{j}\lambda^{n-j}=f(\lambda)\mu^{2}%
\]
and the originating hamiltonians $H_{i}$ constitute a completely integrable
system that we call a system of Benenti type due to S. Benenti's contribution
to the study of these objects \cite{Ben1},\cite{Ben2}. In this case it is
possible to give compact formulas for many objects introduced above. Thus, for
example, the metric $G$ and the Killing tensors $K_{i}$ in (\ref{Stackham})
are given explicitly as%
\[
G=\operatorname*{diag}\left(  \frac{f(\lambda_{1})}{\Delta_{1}},\ldots
,\frac{f(\lambda_{n})}{\Delta_{n}}\right)  ,\text{ \ \ }\Delta_{i}=%
{\textstyle\prod\limits_{j\neq i}}
(\lambda_{i}-\lambda_{j})
\]%
\[
K_{i}=-\operatorname*{diag}\left(  \frac{\partial\rho_{i}}{\partial\lambda
_{1}},\cdots,\frac{\partial\rho_{i}}{\partial\lambda_{n}}\right)  \text{
\ \ \ }i=1,\ldots,n.
\]
Here and below $\rho_{i}=\rho_{i}(\lambda)$ are Vi\`{e}te polynomials (signed
symmetric polynomials) in $\lambda$:%
\begin{equation}
\rho_{i}(\lambda)=(-1)^{i}%
{\displaystyle\sum\limits_{1\leq s_{1}<s_{2}<\ldots<s_{i}\leq n}}
\lambda_{s_{1}}\ldots\lambda_{s_{i}}\text{, \ \ }i=1,\ldots,n \label{defq}%
\end{equation}
that can also be considered as new coordinates on\ the Riemannian manifold
$\mathcal{Q}$ (we will then refer to them as Vi\`{e}te coordinates). Notice
again that the Killing tensors $K_{i}$ do not depend on a particular choice of
$f$ and $\sigma$. It can be shown that as long as $f$ is a polynomial of
degree $\leq n$ then the metric $G$ is flat while if $f$ is a polynomial of
degree $n+1$ then $G$ has constant but non-zero curvature. For the Benenti
class the control matrix $F_{\gamma}$ in (\ref{Fg}) attains the simple form
\begin{equation}
F=\left(
\begin{array}
[c]{cccc}%
-\rho_{1} & 1 &  & \\
-\rho_{2} &  & \ddots & \\
\vdots &  &  & 1\\
-\rho_{n} & 0 & \cdots & 0
\end{array}
\right)  \label{Fmat}%
\end{equation}
and since $V^{(0)}=(0,0,\ldots,0,-1)^{T}$ we easily obtain that the potentials
$V^{(1)}=FV^{(0)}=(0,0,\ldots0,-1,0)^{T}$, $V^{(2)}=F^{2}V^{(0)}%
=(0,0,\ldots0,-1,0,0)^{T}$ up to $V^{(n-1)}=F^{n-1}V^{(0)}=(-1,0,\ldots
,0)^{T}$, are trivial (constant), $V^{(n)}=F^{n}V^{(0)}=(\rho_{1},\ldots
,\rho_{n})$ is the first nontrivial positive potential while $V^{(-1)}%
=F^{-1}V^{(0)}$ $=(1/\rho_{n},\rho_{1}/\rho_{n},\ldots,\rho_{n-1}/\rho
_{n})^{T}$ and so on. More information on Benenti systems can be found in
\cite{bensol,bensol2,bensol3}.

\section{St\"{a}ckel equivalence of St\"{a}ckel systems}

We will now turn to the main question of this article: how to relate two
St\"{a}ckel systems by a single St\"{a}ckel transform and in such a way that
their solutions are related by a reciprocal transform? As we mentioned above,
The Hamiltonians $H_{i}$ defined by (\ref{Stack}) or by (\ref{Stackc}) do not
depend on any additional parameters $\alpha_{i}$ so in order to perform a
St\"{a}ckel transform on (\ref{Stack}) we have to embed it into a
parameter-dependent system. Of course, there is infinitely many ways of
embedding of our St\"{a}ckel system into an $n$-parameter system but the
choice below is natural in the sense that the corresponding St\"{a}ckel
transform transforms a St\"{a}ckel system into a new St\"{a}ckel system. Thus,
consider $n$ Hamiltonians $h_{i}=h_{i}(\lambda,\mu,\alpha)$ defined by the
separation curve%

\begin{equation}
P(\lambda,h,\alpha)\equiv\sigma(\lambda)+%
{\displaystyle\sum\limits_{j=1}^{n}}
h_{j}\lambda^{\gamma_{j}}+R^{-1}(\lambda)%
{\displaystyle\sum\limits_{j=1}^{n}}
\alpha_{j}\lambda^{\delta_{j}}=f(\lambda)\mu^{2} \label{curve1}%
\end{equation}
where $\gamma_{1},\ldots,\gamma_{n}$ and $\delta_{1},\ldots,\delta_{n}$ are
two sequences of integers such that no two $\gamma_{i}$ coincide and similarly
no two $\delta_{i}$ coincide (but we do admit the possibility that some or all
of $\gamma_{i}$ coincide with some $\delta_{i}$) and where $R(\lambda)$ is an
arbitrary meromorphic function of one variable so that%
\[
R(\lambda)=\prod_{s=1}^{k_{1}}(\lambda-\beta_{s})\prod_{s=1}^{k_{2}}%
(\lambda-\beta_{s}^{\prime})^{-1}%
\]
for some (complex in general) constants $\beta_{1},\ldots,\beta_{k_{1}}$ and
$\beta_{1}^{\prime},\ldots,\beta_{k_{2}}^{\prime}$. This function can be
generalized to a matrix function i.e. we define, for any $n\times n$ matrix
$A$ ($\lambda$-dependent or not)
\begin{equation}
R(A)=\prod_{s=1}^{k_{1}}(A-\beta_{s})\prod_{s=1}^{k_{2}}(A-\beta_{s}^{\prime
})^{-1} \label{erekm}%
\end{equation}
(note that all the terms in (\ref{erekm}) commute so that there is no ordering
problem here). The relations (\ref{curve1}) can now be written in a matrix
form as%
\begin{equation}
S_{\gamma}h+R^{-1}(\Lambda)S_{\delta}\alpha=U \label{3}%
\end{equation}
where $S_{\gamma}$ and $S_{\delta}$ are two St\"{a}ckel matrices given by
(\ref{Sg}) (so that $\left(  S_{\gamma}\right)  _{ij}=\lambda_{i}^{\gamma_{j}%
}$ and $\left(  S_{\delta}\right)  _{ij}=\lambda_{i}^{\delta_{j}}$),
\thinspace$h=(h_{1},\ldots,h_{n})^{T}$ is the column vector consisting of
Hamiltonians $h_{i}$, $\alpha=(\alpha_{1},\ldots,\alpha_{n})^{T}$, $U$ is the
column vector given in (\ref{Kvec}) and where $\Lambda=$ diag$(\lambda
_{1},\ldots,\lambda_{n})$ as before. Solving (\ref{3}) with respect to $h$ we
obtain%
\begin{equation}
h=S_{\gamma}^{-1}U-S_{\gamma}^{-1}R^{-1}(\Lambda)S_{\delta}\alpha=S_{\gamma
}^{-1}U-S_{\gamma}^{-1}R^{-1}(\Lambda)S_{\gamma}S_{\gamma}^{-1}S_{\delta
}\alpha. \label{3r}%
\end{equation}

\begin{lemma}
\label{bezbet}In the notation as above%
\[
S_{\gamma}^{-1}R(\Lambda)S_{\gamma}=R(F_{\gamma}).
\]

\end{lemma}

\begin{proof}
We show it for $R(\lambda)=\lambda-\beta$ as the general statement follows
easily by developing the argument below.%
\[
S_{\gamma}^{-1}\left(  \Lambda-\beta\right)  S_{\gamma}=S_{\gamma}^{-1}\Lambda
S_{\gamma}-\beta I=F_{\gamma}-\beta I=R(F_{\gamma}).
\]

\end{proof}

Thus, introducing the shorthand notation%
\[
W_{\delta,\gamma}=-S_{\gamma}^{-1}S_{\delta}%
\]
we see that (\ref{3r}) can be written as%
\begin{equation}
h=H+R^{-1}(F_{\gamma})W_{\delta,\gamma}\alpha\label{3m}%
\end{equation}
where $H=S_{\gamma}^{-1}U$ is the part of $h$ that is independent of
parameters $\alpha_{i}$ (cf. (\ref{lin})). Let us shortly analyze the
structure of the matrix $W_{\delta,\gamma}=-S_{\gamma}^{-1}S_{\delta}$. Assume
that $\sigma(\lambda)$ in (\ref{Vzs}) is a polynomial of the form
$\sigma(\lambda)=\sum_{i=1}^{n}\xi_{i}\lambda^{\delta_{i}}$. Then, as it
follows from (\ref{Vzs}) and from the definition of potentials $V^{(k)}:$%
\[
V^{(\sigma)}=\sum_{i=1}^{n}\xi_{i}V^{(\delta_{i})}.
\]
On the other hand, the formula (\ref{Vzs}) can now be written as%
\[
S_{\gamma}V^{(\sigma)}+S_{\delta}\xi=0
\]
so that $V^{(\sigma)}=-S_{\gamma}^{-1}S_{\delta}\xi=W_{\delta,\gamma}\xi$
which implies that%
\[
\left(  W_{\delta,\gamma}\right)  _{ij}=V_{i}^{(\delta_{j})},
\]
where $V^{(\delta_{j})}=F_{\gamma}^{\delta_{j}}V^{(0)}$ in accordance with
(\ref{Vk}). Therefore, the formula (\ref{3m}) can be written as%
\begin{equation}
h_{i}=H_{i}+\sum_{j,k=1}^{n}\left(  R^{-1}(F_{\gamma})\right)  _{ij}%
V_{j}^{(\delta_{k})}\alpha_{k}\text{, \ \ \ }i=1,\ldots,n. \label{3c}%
\end{equation}

Let us now perform an $n$-parameter St\"{a}ckel transform of the system given
by the curve (\ref{curve1}). Since the number of parameters $\alpha_{i}$ and
the number of Hamiltonians $h_{i}$ are both the same ($=n$) the St\"{a}ckel
transform consists only of part (\ref{alphy}) and is therefore generated by
the relation $h=\widetilde{\alpha}$ (which implies $\widetilde{h}=\alpha$) in
the vector notation as above. We are now in position to formulate the main
theorem of this paper.

\begin{theorem}
\label{mainthm}The $n$-parameter St\"{a}ckel transform generated by
$h=\widetilde{\alpha}$ transforms the set of $n$ Hamiltonians $h$ defined by
(\ref{curve1}) into the following set of Hamiltonians%
\begin{equation}
\widetilde{h}=-W_{\delta,\gamma}^{-1}R(F_{\gamma})H+W_{\delta,\gamma}%
^{-1}R(F_{\gamma})\widetilde{\alpha} \label{htm}%
\end{equation}
(where $\widetilde{h}=(\widetilde{h}_{1},\ldots,\widetilde{h}_{n})^{T}$ and
similarly $\widetilde{\alpha}=(\widetilde{\alpha}_{1},\ldots,\widetilde
{\alpha}_{n})^{T}$) which constitute a new St\"{a}ckel system with the
separation curve of the form%
\begin{equation}
\widetilde{P}(\lambda,\widetilde{h},\widetilde{\alpha})\equiv R(\lambda
)\sigma(\lambda)+R(\lambda)%
{\displaystyle\sum\limits_{j=1}^{n}}
\widetilde{\alpha}_{j}\lambda^{\gamma_{j}}+%
{\displaystyle\sum\limits_{j=1}^{n}}
\widetilde{h}_{j}\lambda^{\delta_{j}}=R(\lambda)f(\lambda)\mu^{2}.
\label{curve2}%
\end{equation}
Moreover, the reciprocal transformation%
\begin{equation}
d\widetilde{t}=-R^{-1}(F_{\delta}^{T})W_{\delta,\gamma}^{T}dt=-W_{\delta
,\gamma}^{T}R^{-1}(F_{\gamma}^{T})dt \label{recip2}%
\end{equation}
transforms $n$-time solutions $x=x(t_{1},\ldots,t_{n},x_{0})$ of the system
(\ref{curve1}) on $M_{\alpha,\widetilde{\alpha}}\ni x_{0}$ into $n$-time
solutions $\widetilde{x}=\widetilde{x}(\widetilde{t}_{1},\ldots,\widetilde
{t}_{n},x_{0})$ of the system (\ref{curve2}) on the same manifold
$M_{\alpha,\widetilde{\alpha}}$.
\end{theorem}

Note that in spite of the fact that we introduced both systems in the
$(\lambda,\mu)$-variables the matrix formulas (\ref{htm}) and (\ref{recip2})
are not tensor and that they are coordinate-free. They can be therefore freely
applied in any coordinate system on $M$, which will be used in the examples
further on.

\begin{proof}
\bigskip Multiplying the curve (\ref{curve1}) by $R(\lambda)$ we obtain%
\[
R(\lambda)\sigma(\lambda)+R(\lambda)%
{\displaystyle\sum\limits_{j=1}^{n}}
h_{j}\lambda^{\gamma_{j}}+%
{\displaystyle\sum\limits_{j=1}^{n}}
\alpha_{j}\lambda^{\delta_{j}}=R(\lambda)f(\lambda)\mu^{2}%
\]
which after the St\"{a}ckel transform $h=\widetilde{\alpha}$ (so that
$\widetilde{h}=\alpha$) obviously attains the form (\ref{curve2}). Let us
therefore show the formula (\ref{htm}). The separation relations implied by
(\ref{curve2}) can be written in matrix form as%
\[
R(\Lambda)S_{\gamma}\widetilde{\alpha}+S_{\delta}\widetilde{h}=R(\Lambda)U
\]
with the column vector $U$ as in (\ref{Kvec}). Solving this with respect to
$\widetilde{h}$ we obtain
\begin{align*}
\widetilde{h}  &  =S_{\delta}^{-1}R(\Lambda)U-S_{\delta}^{-1}R(\Lambda
)S_{\gamma}\widetilde{\alpha}=S_{\delta}^{-1}R(\Lambda)S_{\gamma}H-S_{\delta
}^{-1}R(\Lambda)S_{\gamma}\widetilde{\alpha}\\
&  =\left(  S_{\delta}^{-1}R(\Lambda)S_{\delta}\right)  \left(  S_{\gamma
}^{-1}S_{\delta}\right)  ^{-1}H-\left(  S_{\delta}^{-1}R(\Lambda)S_{\delta
}\right)  \left(  S_{\gamma}^{-1}S_{\delta}\right)  ^{-1}\widetilde{\alpha}\\
&  =-R(F_{\delta})W_{\delta,\gamma}^{-1}H-R(F_{\delta})W_{\delta,\gamma}%
^{-1}\widetilde{\alpha}%
\end{align*}
so the only remaining thing is to show that $R(F_{\delta})W_{\delta,\gamma
}^{-1}=W_{\delta,\gamma}^{-1}R(F_{\gamma})$ which is equivalent to the
statement%
\[
W_{\delta,\gamma}R(F_{\delta})=R(F_{\gamma})W_{\delta,\gamma}%
\]
that can easily be proved in a fashion similar to proof of Lemma \ref{bezbet}.
Finally, the formula (\ref{recip2}) is obtained by inserting (\ref{3m}) into
(\ref{trczas})
\[
\ d\widetilde{t}=-\left(  \frac{\partial h}{\partial\alpha}\right)
^{T}dt=-\left(  R^{-1}(F_{\gamma})W_{\delta,\gamma}\right)  ^{T}dt,
\]
where we use the fact that $R(A)^{T}$\thinspace$=R(A^{T})$.
\end{proof}

Let us also remark that the relations (\ref{htm}) can be explicitly written as
(cf. (\ref{3c})):%
\[
\widetilde{h}_{i}=-\sum_{j,k=1}^{n}R(F_{\delta})_{ij}\widetilde{V_{j}%
}^{(\delta_{k})}H_{k}+\sum_{j,k=1}^{n}R(F_{\delta})_{ij}\widetilde{V_{j}%
}^{(\delta_{k})}\widetilde{\alpha}_{k}\text{, \ \ \ }i=1,\ldots,n
\]
where the potentials $\widetilde{V_{j}}^{(k)}$ are defined by the separation
curve (cf. (\ref{sepV}))%
\begin{equation}
\lambda^{k}+\widetilde{V}_{1}^{(k)}\lambda^{\delta_{1}}+\widetilde{V}%
_{2}^{(k)}\lambda^{\delta_{2}}+\ldots+\widetilde{V}_{n}^{(k)}\lambda
^{\delta_{1}}=0 \label{sepVt}%
\end{equation}
so that%
\[
\widetilde{V}^{(k)}=F_{\delta}^{k}V^{(0)}.
\]
Notice that due to the form of (\ref{sepV}) and (\ref{sepVt}) the potentials
$V^{(k)}$ and $\widetilde{V}^{(k)}$ are related by the St\"{a}ckel transform
$V^{(k)}=\widetilde{\alpha},\widetilde{V}^{(k)}=\alpha$.

Let us also present an alternative way of proving the formula (\ref{recip2}),
directly involving solutions of (\ref{curve1}) and (\ref{curve2}). It follows
from Remark \ref{rozw} and from the above considerations that the multi-time
solutions of the systems (\ref{curve1}) and (\ref{curve2}) on any common level
surface $M_{\alpha,\widetilde{\alpha}}$ attain the form
\begin{equation}
dt=S_{\gamma}^{T}\frac{d\lambda}{\sqrt{f(\lambda)P(\lambda,\widetilde{\alpha
},\alpha)}},\ \ \ \ \ \ \ d\widetilde{t}=S_{\delta}^{T}\frac{d\lambda}%
{\sqrt{R(\lambda)f(\lambda)\widetilde{P}(\lambda,\alpha,\widetilde{\alpha})}}.
\label{sol2}%
\end{equation}
Now, it is easy to see that $\widetilde{P}(\lambda,\alpha,\widetilde{\alpha
})=R(\lambda)P(\lambda,\widetilde{\alpha},\alpha)$ so that, by (\ref{sol2})
and by the fact that $R(\Lambda)$ is symmetric%
\begin{align*}
d\widetilde{t}  &  =S_{\delta}^{T}\frac{d\lambda}{\sqrt{R(\lambda
)f(\lambda)\widetilde{P}(\lambda,\alpha,\widetilde{\alpha})}}=S_{\delta}%
^{T}\frac{d\lambda}{\sqrt{R^{2}(\lambda)f(\lambda)P(\lambda,\widetilde{\alpha
},\alpha)}}=S_{\delta}^{T}R^{-1}(\Lambda)\frac{d\lambda}{\sqrt{f(\lambda
)P(\lambda,\widetilde{\alpha},\alpha)}}=\\
&  =S_{\delta}^{T}R^{-1}(\Lambda)\left(  S_{\gamma}^{T}\right)  ^{-1}%
dt=\left(  S_{\gamma}^{-1}R^{-1}(\Lambda)S_{\delta}\right)  ^{T}dt=\left(
S_{\gamma}^{-1}S_{\delta}S_{\delta}^{-1}R^{-1}(\Lambda)S_{\delta}\right)
^{T}dt=-(W_{\delta,\gamma}R^{-1}(F_{\delta}))^{T}dt
\end{align*}
thus yielding%
\[
d\widetilde{t}=-R^{-1}(F_{\delta}^{T})W_{\delta,\gamma}^{T}dt=-W_{\delta
,\gamma}^{T}R^{-1}(F_{\gamma}^{T})dt,
\]
which is what we wanted to prove.

On the level of St\"{a}ckel transforms Theorem \ref{mainthm} leads to the
following corollary:

\begin{corollary}
Assume that $f_{1}(\lambda)\neq0$ and $f_{2}(\lambda)\neq0$. Any two
St\"{a}ckel systems of the form
\begin{align*}
\sigma_{1}(\lambda)+%
{\displaystyle\sum\limits_{j=1}^{n}}
H_{j}\lambda^{\gamma_{j}}  &  =f_{1}(\lambda)\mu^{2}\\
\sigma_{2}(\lambda)+%
{\displaystyle\sum\limits_{j=1}^{n}}
\widetilde{H}_{j}\lambda^{\delta_{j}}  &  =f_{2}(\lambda)\mu^{2}%
\end{align*}
that satisfy the condition
\begin{equation}
f_{2}(\lambda)\sigma_{1}(\lambda)=\sigma_{2}(\lambda)f_{1}(\lambda)
\label{war}%
\end{equation}
are St\"{a}ckel-related by the single St\"{a}ckel transform%
\begin{equation}
\widetilde{H}=-W_{\delta,\gamma}^{-1}R(F_{\gamma})H \label{trans}%
\end{equation}
with $R(\lambda)=\frac{f_{2}(\lambda)}{f_{1}(\lambda)}$. In particular, any
two geodesic St\"{a}ckel systems (i.e. with $\sigma_{1}=\sigma_{2}=0$) are
connected by the St\"{a}ckel transform (\ref{trans}) with $R(\lambda
)=\frac{f_{2}(\lambda)}{f_{1}(\lambda)}$.
\end{corollary}

\begin{proof}
By Theorem \ref{mainthm}, the St\"{a}ckel transform (\ref{trans}) with
$R(\lambda)=\frac{f_{2}(\lambda)}{f_{1}(\lambda)}$ transforms the first of the
above St\"{a}ckel systems into the second one provided that $\frac
{f_{2}(\lambda)}{f_{1}(\lambda)}=\frac{g_{2}(\lambda)}{g_{l}(\lambda)}$ which
is exactly the condition (\ref{war}). In case of geodesic systems, the
condition (\ref{war}) is always satisfied.
\end{proof}

\begin{proposition}
The condition (\ref{war}) splits all St\"{a}ckel systems of the form
(\ref{Stack}) into equivalence classes since it is an equivalence relation.
\end{proposition}

\begin{proof}
Indeed, if $\frac{f_{2}(\lambda)}{f_{1}(\lambda)}=\frac{g_{2}(\lambda)}%
{g_{1}(\lambda)}=R_{1}(\lambda)$ and $\frac{f_{3}(\lambda)}{f_{2}(\lambda
)}=\frac{g_{3}(\lambda)}{g_{2}(\lambda)}=R_{2}(\lambda)$ then $\frac
{f_{3}(\lambda)}{f_{1}(\lambda)}=\frac{g_{3}(\lambda)}{g_{1}(\lambda)}%
=R_{2}(\lambda)R_{1}(\lambda)$ so this relation is transitive. Further, if
$\frac{f_{2}(\lambda)}{f_{1}(\lambda)}=\frac{g_{2}(\lambda)}{g_{1}(\lambda
)}=R(\lambda)$ then $\frac{f_{1}(\lambda)}{f_{2}(\lambda)}=\frac{g_{1}%
(\lambda)}{g_{2}(\lambda)}=\frac{1}{R(\lambda)}$ so this relation is
reflexive. Finally, $\frac{f_{1}(\lambda)}{f_{1}(\lambda)}=\frac{g_{1}%
(\lambda)}{g_{1}(\lambda)}=1$ so it is a symmetric relation.\newline
\end{proof}

Our formulas contain two special cases: when $\gamma=\delta$ and when $R=1$.
In the first case (i.e. when $\gamma_{i}=\delta_{i},$ $i=1,\ldots,n$; Benenti
systems are in this class) we relate systems belonging to the same class,
where the class is understanding as a fixed sequence $\gamma_{1}%
,...,\gamma_{n},$ and differ by $f$ and $\sigma.$The matrix $W_{\gamma,\gamma
}=-I$ while $F_{\delta}=F_{\gamma}$ so that the formula (\ref{htm}) becomes%
\begin{equation}
\widetilde{h}=R(F_{\gamma})H-R(F_{\gamma})\widetilde{\alpha} \label{ziuta}%
\end{equation}
while (\ref{recip2}) attains the form%
\[
d\widetilde{t}=R^{-1}(F_{\gamma}^{T})dt.
\]
In the second case ($R=1$) we relate systems from different classes, i.e.
$(\gamma_{1},...,\gamma_{n})$ and $(\delta_{1},...,\delta_{n})$ respectively,
which share the same $f$ and $\sigma$. The formula (\ref{htm}) becomes%
\[
\widetilde{h}=-W_{\delta,\gamma}^{-1}H+W_{\delta,\gamma}^{-1}\widetilde
{\alpha}%
\]
while the formula (\ref{recip2}) attains the form%
\begin{equation}
d\widetilde{t}=-W_{\delta,\gamma}^{T}dt. \label{rec}%
\end{equation}

Thus, the general transformation between the systems (\ref{curve1}) and
(\ref{curve2}) can be considered as composition of two transformations: a map
between two St\"{a}ckel systems from the same class (i.e. with $\gamma=\delta
$) but with different $f$ (i.e. metrics) and/or different $\sigma$ and the
transformation between two St\"{a}ckel systems sharing the same $f$ and
$\sigma$ but from different classes. Both these transformations commute.

\section{Examples}

We will now present two examples of our formulas. In order to relate to known
integrable systems we will present both examples in their natural (physical) coordinates.

In our fist example we will relate two families of separation curves for $n=2
$, namely%
\begin{equation}
P(\lambda,h,\alpha)\equiv\sigma(\lambda)+h_{1}\lambda+h_{2}+\lambda(\alpha
_{1}\lambda^{2}+\alpha_{2})=\frac{1}{2}\lambda\mu^{2} \label{k1}%
\end{equation}
and%
\begin{equation}
\widetilde{P}(\lambda,\widetilde{h},\widetilde{\alpha})\equiv\sigma
(\lambda)\lambda^{-1}+\lambda^{-1}(\widetilde{\alpha}_{1}\lambda
+\widetilde{\alpha}_{2})+\widetilde{h}_{1}\lambda^{2}+\widetilde{h}_{2}%
=\frac{1}{2}\mu^{2} \label{k2}%
\end{equation}
which are particular cases of (\ref{curve1}) respectively (\ref{curve2}) with
$(\gamma_{1},\gamma_{2})=(1,0)$, $(\delta_{1},\delta_{2})=(2,0)$ and with
$R=\lambda^{-1},$ while $\sigma(\lambda)$ is for now assumed to be an
arbitrary rational function of $\lambda$. The family (\ref{k1}) contains in
particular a well known Henon-H\`{e}iles (HH) system while the family
(\ref{k2}) contains in particular Drach system \cite{Tsig}. Note that in both
of the above curves one $\gamma_{i}$ coincides with one $\delta_{i} $ and
therefore we can regroup the terms in the above curves to obtain%
\begin{equation}
\sigma(\lambda)+\alpha_{1}\lambda^{3}+(h_{1}+\alpha_{2})\lambda+h_{2}=\frac
{1}{2}\lambda\mu^{2} \label{HH}%
\end{equation}
for the HH family (\ref{k1}) and%
\begin{equation}
\sigma(\lambda)\lambda^{-1}+\widetilde{h}_{1}\lambda^{2}+(\widetilde{h}%
_{2}+\widetilde{\alpha}_{1})+\widetilde{\alpha}_{2}\lambda^{-1}=\frac{1}{2}%
\mu^{2} \label{Drach}%
\end{equation}
for the Drach family (\ref{k2}). Since both metrics are flat we will now
construct both systems in their respective flat coordinates. As an
intermediate step, we will write the systems in Vi\`{e}te coordinates
(\ref{defq}) that now attain the form%
\begin{equation}
\rho_{1}=-\lambda_{1}-\lambda_{2}\text{, \ \ }\rho_{2}=\lambda_{1}\lambda_{2}.
\label{V2}%
\end{equation}

In the above coordinates the control matrix $F_{\gamma}$ (cf. (\ref{Fg}) and
(\ref{Fmat})) of the HH family is%
\[
F_{\gamma}=\left(
\begin{array}
[c]{cc}%
-\rho_{1} & 1\\
-\rho_{2} & 0
\end{array}
\right)
\]
and as we remember it is coordinate free. The passage to the flat coordinates
$(x_{1},x_{2})$ is given by the point transformation \cite{macartur2007}
\[
\rho_{1}=x_{1},\text{ \ }\rho_{2}=-\frac{1}{4}x_{2}^{2}.
\]
The metric $G$ and the Killing tensor $K_{2}$ are now%
\[
G=\left(
\begin{array}
[c]{cc}%
1 & 0\\
0 & 1
\end{array}
\right)  \text{, \ }K_{2}=\left(
\begin{array}
[c]{cc}%
0 & -\frac{1}{2}x_{2}\\
-\frac{1}{2}x_{2} & x_{1}%
\end{array}
\right)
\]
so that geodesic (i.e. with $\sigma(\lambda)=0$) Hamiltonians of (\ref{HH})
are%
\[
E_{1}=\frac{1}{2}y_{1}^{2}+\frac{1}{2}y_{2}^{2}\text{, \ \ \ \ \ }E_{2}%
=\frac{1}{2}x_{1}y_{2}^{2}-\frac{1}{2}x_{2}y_{1}y_{2},
\]
where $(y_{1},y_{2})^{T}$ are momenta conjugate to $(x_{1},x_{2})$. \ Note
that $(x_{1},x_{2})$ are not only flat but also orthogonal coordinates for
(\ref{HH}). The matrix $F_{\gamma}$ is now%
\[
F_{\gamma}=\left(
\begin{array}
[c]{cc}%
-x_{1} & 1\\
\frac{1}{4}x_{2}^{2} & 0
\end{array}
\right)
\]
so that the Hamiltonians of (\ref{HH}) in the flat coordinates are%
\begin{align*}
h_{1}  &  =E_{1}+V_{1}^{(\sigma)}-\left(  x_{1}^{2}+\frac{1}{4}x_{2}%
^{2}\right)  \alpha_{1}-\alpha_{2}\\
h_{1}  &  =E_{1}+V_{2}^{(\sigma)}+\frac{1}{4}x_{1}x_{2}^{2}\alpha_{1},
\end{align*}
where due to (\ref{Vsigma}) we have $V^{(\sigma)}=\sigma(F_{\gamma}%
)V^{(0)}=\sigma(F_{\gamma})(0,-1)^{T}$. Let us now confine ourselves to the
following three-parameter set of separable potentials:%
\begin{equation}
\sigma(\lambda)=b_{1}\lambda^{5}+b_{2}\lambda^{4}+b_{3}\lambda^{2},\text{
\ }b_{i}\in\mathbf{R.} \label{specs}%
\end{equation}
We can now calculate $V^{(\sigma)}$ explicitly and we find%
\begin{align*}
h_{1}  &  =E_{1}-b_{1}\left(  x_{1}^{4}+\frac{3}{4}x_{1}^{2}x_{2}^{2}+\frac
{1}{16}x_{2}^{4}\right)  +b_{2}\left(  x_{1}^{3}+\frac{1}{2}x_{1}x_{2}%
^{2}\right)  +b_{3}x_{1}-\left(  x_{1}^{2}+\frac{1}{4}x_{2}^{2}\right)
\alpha_{1}-\alpha_{2}\\
h_{2}  &  =E_{2}+b_{1}\left(  \frac{1}{4}x_{1}^{3}x_{2}^{2}+\frac{1}{8}%
x_{1}x_{2}^{4}\right)  -b_{2}\left(  \frac{1}{4}x_{1}^{2}x_{2}^{2}+\frac
{1}{16}x_{2}^{3}\right)  -\frac{1}{4}b_{3}x_{2}^{2}+\frac{1}{4}x_{1}x_{2}%
^{2}\alpha_{1}%
\end{align*}
and for $\alpha_{1}=\alpha_{2}=0$, $b_{1}=b_{3}=0,b_{2}=1$ we receive the
classical H\`{e}non-Heiles system.

Consider now the Drach family (\ref{Drach}). The matrix $F_{\delta}$ in
Vi\`{e}te coordinates (\ref{V2}) has the form%
\[
F_{\delta}=\left(
\begin{array}
[c]{cc}%
\frac{-\rho_{1}^{2}+\rho_{2}}{\rho_{1}} & -\frac{1}{\rho_{1}}\\
\frac{\rho_{2}^{2}}{\rho_{1}} & -\frac{\rho_{2}}{\rho_{1}}%
\end{array}
\right)  .
\]
We pass now to flat but non-orthogonal coordinates $(x,y\,)$ given by
\cite{Tsig}%
\[
\rho_{1}=-2x^{\frac{1}{2}}\text{, \ }\rho_{2}=x-y.
\]
The metric $\widetilde{G}$ and the Killing tensor $\widetilde{K}_{2}$ of this
system are%
\[
\widetilde{G}=\left(
\begin{array}
[c]{cc}%
0 & \frac{1}{2}\\
\frac{1}{2} & 0
\end{array}
\right)  \text{, \ }\widetilde{K}_{2}=\left(
\begin{array}
[c]{cc}%
-(x+y) & 2x\\
2y & -(x+y)
\end{array}
\right)
\]
so that the geodesic Hamiltonians of (\ref{Drach}) are%
\[
\widetilde{E}_{1}=\frac{1}{2}p_{x}p_{y}\text{, \ \ \ \ \ }\widetilde{E}%
_{2}=\frac{1}{2}xp_{x}^{2}+\frac{1}{2}yp_{y}^{2}-\frac{1}{2}(x+y)p_{x}p_{y},
\]
where $(p_{x},p_{y})$ are momenta conjugate to $(x,y)$. The matrix $F_{\delta
}$ becomes%
\[
F_{\delta}=\left(
\begin{array}
[c]{cc}%
\frac{1}{2}x^{-\frac{1}{2}}(3x+y) & \frac{1}{2}x^{-\frac{1}{2}}\\
-\frac{1}{2}x^{-\frac{1}{2}}(x-y)^{2} & \frac{1}{2}x^{-\frac{1}{2}}(x-y)
\end{array}
\right)
\]
and hence%
\begin{align*}
\widetilde{h}_{1}  &  =\widetilde{E}_{1}+\widetilde{V}_{1}^{(\sigma)}+\frac
{1}{2}x^{-\frac{1}{2}}(x-y)^{-1}\widetilde{\alpha}_{2}\\
\widetilde{h}_{2}  &  =\widetilde{E}_{2}+\widetilde{V}_{2}^{(\sigma
)}-\widetilde{\alpha}_{1}-\frac{1}{2}x^{-\frac{1}{2}}(x-y)^{-1}%
(3x+y)\widetilde{\alpha}_{2}%
\end{align*}
where due to (\ref{Vsigma}) and to (\ref{Drach}) we have $\widetilde
{V}^{(\sigma)}=\sigma(F_{\delta})F_{\delta}^{-1}\widetilde{V}^{(0)}%
=\sigma(F_{\delta})F_{\delta}^{-1}(0,-1)^{T}$. For our particular choice of
$\sigma(\lambda)$ as in (\ref{specs}) we have $\sigma(\lambda)\lambda
^{-1}=b_{1}\lambda^{4}+b_{2}\lambda^{3}+b_{3}\lambda$ and the Hamiltonians
$\widetilde{h}_{i}$ attain the explicit form%
\begin{align*}
\widetilde{h}_{1}  &  =\widetilde{E}_{1}-2b_{1}(x+y)-\frac{1}{2}b_{2}%
x^{-\frac{1}{2}}(3x+y)-\frac{1}{2}b_{3}x^{-\frac{1}{2}}+\frac{1}{2}%
x^{-\frac{1}{2}}(x-y)^{-1}\widetilde{\alpha}_{2}\\
\widetilde{h}_{2}  &  =\widetilde{E}_{2}+b_{1}(x-y)^{2}+\frac{1}{2}%
b_{2}x^{-\frac{1}{2}}(x-y)^{2}-\frac{1}{2}b_{3}x^{-\frac{1}{2}}%
(x-y)-\widetilde{\alpha}_{1}-\frac{1}{2}x^{-\frac{1}{2}}(x-y)^{-1}%
(3x+y)\widetilde{\alpha}_{2}.
\end{align*}
The above Hamiltonians become, after identification of constants
$-2b_{1}=\alpha,$ $-\frac{1}{2}b_{3}=\beta$, $-\frac{1}{2}b_{2}=\gamma$,
$\widetilde{\alpha}_{2}=0$ identical with the three-parameter Drach systems
given in \cite{Tsig}.

Let us now perform the transform between both families, i.e. between
(\ref{HH}) and (\ref{Drach}). According to formula (\ref{htm}) in Theorem
\ref{mainthm} the parameter independent parts of Hamiltonians transform as%
\[
\widetilde{H}=-W_{\delta,\gamma}^{-1}R(F_{\gamma})H\equiv CH
\]
with $R(F_{\gamma})=F_{\gamma}^{-1}$ and with%
\[
W_{\delta,\gamma}=\left(
\begin{array}
[c]{cc}%
V_{1}^{(2)} & 0\\
V_{2}^{(2)} & -1
\end{array}
\right)  =\left(
\begin{array}
[c]{cc}%
x_{1} & 0\\
-\frac{1}{4}x_{2}^{2} & -1
\end{array}
\right)  ,
\]
so that%
\[
C=\left(
\begin{array}
[c]{cc}%
0 & -\frac{4}{x_{1}x_{2}^{2}}\\
1 & \frac{4x_{1}^{2}+x_{2}^{2}}{x_{1}x_{2}^{2}}%
\end{array}
\right)  .
\]
Moreover, the map (\ref{recip2}) between solutions (\ref{sol2}) of (\ref{HH})
and (\ref{Drach}) on $M_{a,\widetilde{\alpha}}$ becomes%
\[
d\widetilde{t}=-W_{\delta,\gamma}^{T}R^{-1}(F_{\gamma}^{T})dt=A^{T}dt
\]
where%
\[
A=C^{-1}=-\frac{\partial h}{\partial\alpha}=\left(
\begin{array}
[c]{cc}%
x_{1}^{2}+\frac{1}{4}x_{2}^{2} & 1\\
-\frac{1}{4}x_{1}x_{2}^{2} & 0
\end{array}
\right)
\]
and this map leaves the common level surface $M_{\alpha,\widetilde{\alpha}}$
invariant. Note also that now $\widetilde{P}(\lambda,\alpha_{1},\alpha
_{2},\widetilde{\alpha}_{1},\widetilde{\alpha}_{2})=\lambda^{-1}%
P(\lambda,\widetilde{\alpha}_{1},\widetilde{\alpha}_{2},\alpha_{1},\alpha
_{2})$. The point transformation between flat coordinates of both families is
\[
x_{1}=-2x^{1/2},\ \ \ x_{2}=2(y-x)^{1/2}\ \Longrightarrow\ x=\frac{1}{2}%
x_{1}^{2},\ \ y=\frac{1}{4}x_{1}^{2}+\frac{1}{4}x_{2}^{2}.
\]

In our second example we will relate the HH family of separable potentials
with a family of elliptic separable potentials. Both systems will belong to
the same class of St\"{a}ckel systems i.e. $\gamma_{i}=\delta_{i}$ for $i=1,2$
now (so that $W_{\gamma,\delta}=-I$ and $F_{\gamma}=F_{\delta}$ which is valid
in any coordinate system) but they have different metrics. Let us thus first
recollect some basic fact about generalized elliptic coordinates and a
hierarchy of elliptic separable potentials \cite{stefan}. Denote by
$(q_{1},...,q_{n})$ the Euclidian coordinates on $\mathbf{R}^{n}$ and by
$(p_{1},...,p_{n})$ the conjugate momenta. The generalized Jacobi elliptic
coordinates $(\lambda_{1},...,\lambda_{n})$ are defined by
\begin{equation}
1+\frac{1}{4}\sum_{k=1}^{n}\frac{q_{k}^{2}}{(z-\beta_{k})}=\frac{\prod
_{j=1}^{n}(z-\lambda_{j})}{\prod_{j=1}^{n}(z-\beta_{j})}, \label{j}%
\end{equation}
where $\beta_{i}$ are nonzero different constants. Let us introduce the
following abbreviations
\begin{equation}
B(z)=\prod_{j=1}^{n}(z-\beta_{j}),\ \ \ \Lambda(z)=\prod_{j=1}^{n}%
(z-\lambda_{j}),\ \ \ \frac{B(z)}{(z-\beta_{k})}=B_{k}(z)=-\sum_{j=1}^{n}%
\frac{\partial\rho_{j}(\beta)}{\partial\beta_{k}}z^{n-j}, \label{j1}%
\end{equation}
where $\rho_{j}$ are Vi\`{e}te polynomials with respect to its arguments. Then
(\ref{j}) takes the form
\begin{equation}
B(z)+\frac{1}{4}\sum_{k=1}^{n}B_{k}(z)q_{k}^{2}=\Lambda(z). \label{j2}%
\end{equation}
For $z=\beta_{i}:~B(\beta_{i})=0,\ B_{k}(\beta_{i})=\delta_{ki}B_{k}(\beta
_{k}),$ hence
\[
q_{k}^{2}=4\frac{\Lambda(\beta_{k})}{B_{k}(\beta_{k})}=4\frac{\prod_{j=1}%
^{n}(\beta_{k}-\lambda_{j})}{\prod_{j=1,j\neq k}^{n}(\beta_{k}-\beta_{j})}%
\]
and then from (\ref{j1}) and (\ref{j2})
\begin{equation}
\rho_{j}(\lambda)=\rho_{j}(\beta)-\frac{1}{4}\sum_{j=1}^{n}\frac{\partial
\rho_{j}(\beta)}{\partial\beta_{k}}q_{k}^{2}. \label{p5}%
\end{equation}
Elliptic coordinates are separation coordinates for the following family of
Benenti systems
\[
\sigma(\lambda)+H_{1}\lambda^{n-1}+...+H_{n}=-\frac{1}{2}B(\lambda)\mu^{2},
\]
where the family of elliptic separable potentials is given by $\sigma
(\lambda)=\lambda^{k},$ $k\in\mathbb{Z}$, and thus $V^{(k)}(q)=F^{k}%
(q)V^{(0)},$ where $F$ is given by (\ref{Fg}) and (\ref{p5}).

Let us relate two separation curves for $n=2$, namely%
\begin{equation}
P(\lambda,h,\alpha)\equiv\sigma(\lambda)+h_{1}\lambda+h_{2}-B^{-1}%
(\lambda)\lambda(\alpha_{1}\lambda+\alpha_{2})=\frac{1}{2}\lambda\mu^{2}
\label{p7}%
\end{equation}
and%
\begin{equation}
\widetilde{P}(\lambda,\widetilde{h},\widetilde{\alpha})\equiv-B(\lambda
)\sigma(\lambda)\lambda^{-1}-B(\lambda)(\widetilde{\alpha}_{1}+\widetilde
{\alpha}_{2}\lambda^{-1})+\widetilde{h}_{1}\lambda+\widetilde{h}_{2}=-\frac
{1}{2}B(\lambda)\mu^{2} \label{p8}%
\end{equation}
where both families are now from Benenti class with $(\gamma_{1},\gamma
_{2})=(\delta_{1},\delta_{2})=(1,0)$ and with $R(\lambda)=-B(\lambda
)\lambda^{-1}=-(\lambda-\beta_{1})(\lambda-\beta_{2})\lambda^{-1},$ while
$\sigma(\lambda)$ is again assumed to be an arbitrary rational function of
$\lambda$. For the extended H\`{e}non-Heiles system, when $\sigma
(\lambda)=\lambda^{4}$ in (\ref{p7}), in flat orthogonal coordinates from the
previous example, an appropriate Hamiltonians are
\begin{align*}
h_{1}  &  =\frac{1}{2}y_{1}^{2}+\frac{1}{2}y_{2}^{2}+x_{1}^{3}+\frac{1}%
{2}x_{1}x_{2}^{2}+\frac{4(\beta_{1}+\beta_{2})x_{2}^{2}-16\beta_{1}\beta
_{2}x_{1}}{(4\beta_{1}^{2}+4\beta_{1}x_{1}-x_{2}^{2})(4\beta_{2}^{2}%
+4\beta_{2}x_{1}-x_{2}^{2})}\alpha_{1}\\
&  +\frac{4x_{2}^{2}+16\beta_{1}\beta_{2}}{(4\beta_{1}^{2}+4\beta_{1}%
x_{1}-x_{2}^{2})(4\beta_{2}^{2}+4\beta_{2}x_{1}-x_{2}^{2})}\alpha_{2},
\end{align*}%
\begin{align*}
h_{2}  &  =\frac{1}{2}x_{1}y_{2}^{2}-\frac{1}{2}x_{2}y_{1}y_{2}-\frac{1}%
{4}x_{1}^{2}x_{2}^{2}-\frac{1}{16}x_{2}^{4}+\frac{x_{2}^{4}+4\beta_{1}%
\beta_{2}x_{2}^{2}}{(4\beta_{1}^{2}+4\beta_{1}x_{1}-x_{2}^{2})(4\beta_{2}%
^{2}+4\beta_{2}x_{1}-x_{2}^{2})}\alpha_{1}\\
&  +\frac{4x_{1}x_{2}^{2}+4(\beta_{1}+\beta_{2})x_{2}^{2}}{(4\beta_{1}%
^{2}+4\beta_{1}x_{1}-x_{2}^{2})(4\beta_{2}^{2}+4\beta_{2}x_{1}-x_{2}^{2}%
)}\alpha_{2}.
\end{align*}

We pass now to the respective system from the family (\ref{p8}). The
transformation from Vi\`{e}te to flat orthogonal coordinates is given by
(\ref{p5})%
\[
\rho_{1}(\lambda)=-\beta_{1}-\beta_{2}+\frac{1}{4}q_{1}^{2}+\frac{1}{4}%
q_{2}^{2}\text{, \ }\rho_{2}(\lambda)=\beta_{1}\beta_{2}-\frac{1}{4}\beta
_{2}q_{1}^{2}-\frac{1}{4}\beta_{1}q_{2}^{2}.
\]
The metric $\widetilde{G}$ and the Killing tensor $\widetilde{K}_{2}$ of this
system are%
\[
\widetilde{G}=\left(
\begin{array}
[c]{cc}%
1 & 0\\
0 & 1
\end{array}
\right)  \text{, \ }\widetilde{K}_{2}=\left(
\begin{array}
[c]{cc}%
-\beta_{2}+\frac{1}{4}q_{2}^{2} & -\frac{1}{4}q_{1}q_{2}\\
-\frac{1}{4}q_{1}q_{2} & -\beta_{1}+\frac{1}{4}q_{1}^{2}%
\end{array}
\right)
\]
so that the geodesic Hamiltonians of (\ref{p8}) are%
\[
\widetilde{E}_{1}=\frac{1}{2}p_{1}^{2}+\frac{1}{2}p_{2}^{2}\text{,
\ \ \ \ \ }\widetilde{E}_{2}=\frac{1}{2}(-\beta_{2}+\frac{1}{4}q_{2}^{2}%
)p_{1}^{2}+\frac{1}{2}(-\beta_{1}+\frac{1}{4}q_{1}^{2})p_{2}^{2}-\frac{1}%
{4}q_{1}q_{2}p_{1}p_{2}.
\]
The matrix $\widetilde{F}_{\gamma}$ becomes%
\[
\widetilde{F}_{\gamma}=\left(
\begin{array}
[c]{cc}%
\beta_{1}+\beta_{2}-\frac{1}{4}q_{1}^{2}-\frac{1}{4}q_{2}^{2} & 1\\
-\beta_{1}\beta_{2}+\frac{1}{4}\beta_{2}q_{1}^{2}+\frac{1}{4}\beta_{1}%
q_{2}^{2} & 0
\end{array}
\right)
\]
and hence%
\[
\widetilde{h}_{1}=\widetilde{E}_{1}+\widetilde{V}_{1}^{(\sigma)}-\frac{1}%
{4}(q_{1}^{2}+q_{2}^{2})\widetilde{\alpha}_{1}-\frac{\beta_{2}q_{1}^{2}%
+\beta_{1}q_{2}^{2}}{4\beta_{1}\beta_{2}-\beta_{2}q_{1}^{2}-\beta_{1}q_{2}%
^{2}}\widetilde{\alpha}_{2},
\]%
\[
\widetilde{h}_{2}=\widetilde{E}_{2}+\widetilde{V}_{2}^{(\sigma)}+\frac{1}%
{4}(\beta_{2}q_{1}^{2}+\beta_{1}q_{2}^{2})\widetilde{\alpha}_{1}+\frac
{\beta_{2}^{2}q_{1}^{2}+\beta_{1}^{2}q_{2}^{2}}{4\beta_{1}\beta_{2}-\beta
_{2}q_{1}^{2}-\beta_{1}q_{2}^{2}}\widetilde{\alpha}_{2},
\]
where $\widetilde{V}^{(\sigma)}=-B(F_{\gamma})F_{\gamma}^{3}V^{(0)},$ so
\begin{align*}
\widetilde{V}_{1}^{(\sigma)}  &  =-\frac{1}{4}(\beta_{1}^{3}q_{1}^{2}%
+\beta_{2}^{3}q_{2}^{2})+\frac{1}{8}(\beta_{1}\beta_{2}+\beta_{1}^{2}%
+\beta_{2}^{2})q_{1}^{2}q_{2}^{2}+\frac{3}{16}(\beta_{1}^{2}q_{1}^{4}%
+\beta_{2}^{2}q_{2}^{4})-\frac{3}{64}(2\beta_{1}+\beta_{2})q_{1}^{4}q_{2}%
^{2}\\
&  -\frac{3}{64}(\beta_{1}+2\beta_{2})q_{1}^{2}q_{2}^{4}-\frac{3}{64}%
(\beta_{1}q_{1}^{6}+\beta_{2}q_{2}^{6})+\frac{1}{64}(q_{1}^{6}q_{2}^{2}%
+q_{1}^{2}q_{2}^{6})+\frac{3}{128}q_{1}^{4}q_{2}^{4}+\frac{1}{256}(q_{1}%
^{8}+q_{2}^{8}),
\end{align*}%
\begin{align*}
\widetilde{V}_{2}^{(\sigma)}  &  =\frac{1}{4}\beta_{1}\beta_{2}(\beta_{2}%
^{2}q_{1}^{2}+\beta_{1}^{2}q_{2}^{2})-\frac{1}{16}(\beta_{1}^{3}+2\beta
_{1}\beta_{2}^{2}+2\beta_{1}\beta_{2}^{2}+\beta_{2}^{3})q_{1}^{2}q_{2}%
^{2}-\frac{3}{16}\beta_{1}\beta_{2}(\beta_{1}q_{1}^{4}+\beta_{2}q_{2}^{4})\\
&  +\frac{1}{32}(\beta_{1}^{2}+\frac{5}{2}\beta_{1}\beta_{2}+\beta_{2}%
^{2})(q_{1}^{4}q_{2}^{2}+q_{1}^{2}q_{2}^{4})\frac{3}{64}\beta_{1}\beta
_{2}(q_{1}^{6}+q_{2}^{6})-\frac{1}{256}(3\beta_{1}+\beta_{2})q_{1}^{2}%
q_{2}^{6}\\
&  -\frac{1}{256}(\beta_{1}+3\beta_{2})q_{1}^{6}q_{2}^{2}-\frac{3}{256}%
(\beta_{1}+\beta_{2})q_{1}^{4}q_{2}^{4}-\frac{1}{256}(\beta_{2}q_{1}^{8}%
+\beta_{1}q_{2}^{8}).
\end{align*}
According with (\ref{ziuta}), we have now
\[
\widetilde{H}=R(F_{\gamma})H=-B(F_{\gamma})F_{\gamma}^{-1}H,
\]
where%
\[
R(F_{\gamma})=\left(
\begin{array}
[c]{cc}%
\beta_{1}+\beta_{2}+x_{1} & -1-4\beta_{1}\beta_{2}x_{2}^{-2}\\
-\frac{1}{4}x_{2}^{2}-\beta_{1}\beta_{2} & \beta_{1}+\beta_{2}-4\beta_{1}%
\beta_{2}x_{1}x_{2}^{-2}%
\end{array}
\right)
\]
while the reciprocal transformation (\ref{rec}) on $M_{a,\widetilde{\alpha}}$
attains the form
\begin{equation}
dt=R(\widetilde{F}_{\gamma}^{T})d\widetilde{t}=-\left(  \frac{\partial
\widetilde{h}}{\partial\widetilde{\alpha}}\right)  ^{T}d\widetilde{t}
\label{rec1}%
\end{equation}
where%

\[
R(\widetilde{F}_{\gamma}^{T})=\left(
\begin{array}
[c]{cc}%
\frac{1}{4}(q_{1}^{2}+q_{2}^{2}) & -\frac{1}{4}(\beta_{2}q_{1}^{2}+\beta
_{1}q_{2}^{2})\\
(\beta_{2}q_{1}^{2}+\beta_{1}q_{2}^{2})(4\beta_{1}\beta_{2}-\beta_{2}q_{1}%
^{2}-\beta_{1}q_{2}^{2})^{-1} & -(\beta_{2}^{2}q_{1}^{2}+\beta_{1}^{2}%
q_{2}^{2})(4\beta_{1}\beta_{2}-\beta_{2}q_{1}^{2}-\beta_{1}q_{2}^{2})^{-1}%
\end{array}
\right)  .
\]
In (\ref{rec1}) we used the inverse of formula (\ref{rec}) to avoid the
complicated matrix $R^{-1}(F_{\gamma}).$ As before the transformation leaves
the common level surface $M_{\alpha,\widetilde{\alpha}}$ invariant. Also, this
time we have $\widetilde{P}(\lambda,\alpha_{1},\alpha_{2},a_{1},a_{2}%
)=-B(\lambda)\lambda^{-1}P(\lambda,a_{1},a_{2},\alpha_{1},\alpha_{2})$. The
point transformation between flat coordinates of both systems is
\[
x_{1}=-\beta_{1}-\beta_{2}+\frac{1}{4}q_{1}^{2}+\frac{1}{4}q_{2}%
^{2},\ \ \ \ x_{2}^{2}=\beta_{1}q_{2}^{2}+\beta_{2}q_{1}^{2}-4\beta_{1}%
\beta_{2}.
\]

\section{Acknowledgement}

Both authors were partially supported by Swedish Research Council grant no VR 624-2011-52.

\end{document}